\documentclass[a4paper]{article}
\usepackage[utf8]{inputenc}
\usepackage{xcolor}
\usepackage{amsfonts}
\usepackage{longtable} 
\usepackage{array} 
\usepackage{amsmath}
\usepackage{amssymb}
\usepackage{authblk} 
\usepackage{multirow} 
\usepackage{lineno}
\usepackage[numbers,sort&compress]{natbib}
\usepackage{graphicx}
\usepackage{subcaption}
\usepackage{doi}
\usepackage{enumerate}
\DeclareUnicodeCharacter{2060}{\nolinebreak}

\usepackage[utf8]{inputenc}
\usepackage[T1]{fontenc}   
\usepackage{hyperref}     
\usepackage{tabularx} 
\usepackage{booktabs}
\hypersetup{      
    colorlinks=true,                
    breaklinks=true,                
    urlcolor= blue,                
    linkcolor= blue,                
    bookmarksopen=false,
    filecolor=black,
    citecolor=blue,
    linkbordercolor=red}
\AtBeginDocument{\hypersetup{pdfborder={0 0 0}}}   
\usepackage{setspace}
\usepackage{geometry}
\title{Non-Unitarity Effects and Fake CP Violation in Neutrino Oscillation Experiments}
\author[$1$]{Pratima Singh}
\author[$1$]{Alina Naqvi}
\author[$1$]{Suhani Yadav}
\author[$1$]{R.B.Singh \thanks{{rajendrasinghrb@gmail.com}}}
\author[$1$]{Jyotsna Singh \thanks{{singh.jyotsnalu@gmail.com}}}

\affil[$1$]{\textit{\small{Department of Physics, University of Lucknow, Lucknow, Uttar Pradesh, 226007, India}}}
\date{}

\begin{document}
\maketitle
\begin{abstract}
Future long-baseline neutrino oscillation experiments aim to establish leptonic CP violation and determine the neutrino mass ordering with unprecedented precision. However, these measurements can be significantly affected by possible deviations from the unitarity of the PMNS mixing matrix, which introduce additional CP-violating phases capable of generating fake CP-violating signals.  We investigate the impact of non-unitary leptonic mixing on CP-violation and mass-ordering measurements at DUNE and Hyper-Kamiokande using GLoBES simulations with a custom non-unitary probability engine. We analyze the energy dependence of the neutrino--antineutrino CP asymmetry, quantify the fake-to-genuine CP asymmetry ratio, evaluate the CP violation discovery sensitivity, and study the hierarchy--CP--non-unitarity degeneracies in the $(\delta_{CP},\phi_{21})$ parameter space. We demonstrate that non-unitary mixing can generate sizeable CP asymmetries even for CP-conserving values of the standard Dirac phase, thereby mimicking genuine leptonic CP violation. While the fake contribution remains below $5\%$ of the genuine signal near each experiment's oscillation maximum, it exceeds the genuine signal in specific intermediate energy windows ($\sim130\%$ at $1.5$--$1.6$~GeV), demonstrating that fake CP violation can dominate over the genuine contribution in these regions. The combined DUNE and Hyper-Kamiokande analysis reduces the allowed $(\delta_{CP},\varphi_{21})$ parameter space by a factor of $7$ at $1\sigma$ and $\sim16$ at $2\sigma$--$3\sigma$, substantially suppressing the degeneracy that neither experiment resolves individually and providing a robust strategy for distinguishing genuine from fake CP violation while improving sensitivity to the neutrino mass ordering.

\end{abstract}

\section{Introduction}
The discovery of neutrino oscillations, recognized by the 2015 Nobel Prize in Physics, established that neutrinos undergo flavor transitions during propagation and therefore possess non-zero masses \cite{Kajita:2010zz, Super-Kamiokande:1998kpq, Bellerive:2016byv, Gonzalez-Garcia:2002bkq,heeger2005evidence,strassler202611,king2013neutrino}. This landmark discovery provided the first direct experimental evidence for physics beyond the minimal Standard Model, in which neutrinos are assumed to be massless. Over the past two decades, remarkable progress has been achieved in the measurement of neutrino oscillation parameters through solar, atmospheric, reactor, and accelerator neutrino experiments \cite{bellini2014neutrino,Capozzi:2025wyn,ParticleDataGroup:2024cfk,Capozzi:2013psa}. Within the standard three-flavor framework, neutrino flavor mixing is described by the Pontecorvo--Maki--Nakagawa--Sakata (PMNS) matrix, which is parameterized by three leptonic mixing angles, $\theta_{12}$, $\theta_{13}$, and $\theta_{23}$, two independent mass-squared differences, $\Delta m_{21}^{2}$ and $\Delta m_{31}^{2}$, and a single Dirac CP-violating phase, $\delta_{CP}$, responsible for CP violation in neutrino oscillations \cite{iizuka2015parametrization, nakamura2018neutrino}. Global analyses have measured the mixing angles and mass-squared differences with remarkable precision, firmly establishing the three-flavor oscillation paradigm as the standard description of neutrino flavor conversion. Nevertheless, several fundamental questions remain unanswered, including the determination of the neutrino mass ordering, the resolution of the $\theta_{23}$ octant ambiguity, the precise measurement of the leptonic Dirac CP phase, and the possible existence of physics beyond the standard three-neutrino framework \cite{kolupaeva2023neutrino,Messier:2016ugx,Branco:2011zb,NOvA:2026tir}.
\\
Among these outstanding problems, the determination of signatures of leptonic CP violation and the neutrino mass ordering constitutes one of the primary scientific objectives of current and future long-baseline neutrino oscillation experiments \cite{ohlsson2013probing}. Considerable theoretical and phenomenological efforts have been devoted to understanding the sensitivity of future experiments to these observables and to investigating the impact of parameter degeneracies, matter effects, and possible new-physics scenarios on their determination \cite{Nagu:2021irq,Naaz:2018amr}. A non-zero value of the Dirac phase $\delta_{CP}$ would manifest itself through a measurable difference between the oscillation probabilities of neutrinos and antineutrinos, thereby establishing intrinsic CP violation in the lepton sector. Although the CP violation responsible for neutrino oscillations is not directly identical to that generating the cosmological baryon asymmetry, nevertheless, the observation of leptonic CP violation would represent a major milestone in particle physics and could provide an essential ingredient for understanding the origin of the observed matter--antimatter asymmetry of the Universe through leptogenesis \cite{faure2024neutrinos}.
\\
Simultaneously, determining the sign of $\Delta m_{31}^{2}$, commonly referred to as the neutrino mass ordering, is crucial for understanding the neutrino mass spectrum, interpreting neutrinoless double-beta decay experiments, constraining models of neutrino mass generation, and improving the precision of future oscillation measurements. Future long-baseline facilities such as the Deep Underground Neutrino Experiment (DUNE) \cite{abi2020volume,DUNE:2020jqi,abi2020deep,DUNE:2015lol} and Hyper-Kamiokande (Hyper-K) \cite{hyper2015physics,Hyper-Kamiokande:2018ofw,Hyper-Kamiokande:2016srs} have been specifically designed to address these outstanding questions with unprecedented precision. Their complementary baselines, neutrino energies, detector technologies, and sensitivities to terrestrial matter effects make them ideally suited for determining the neutrino mass ordering, discovering leptonic CP violation, and probing possible deviations from the standard three-flavor paradigm \cite{Bora:2025xfj}.
\\
The interpretation of future precision measurements, however, may be affected by the presence of new physics beyond the standard three-neutrino paradigm \cite{capozzi2021unfinished}. Several theoretically well-motivated extensions of the Standard Model predict modifications to neutrino oscillations that can introduce additional sources of flavor mixing and CP violation \cite{Medhi:2021wxj, Denton:2020uda}. Such effects may alter the oscillation probabilities measured at long-baseline experiments and generate visible CP-violating signatures that are not associated with the standard Dirac CP phase. Consequently, the observation of a CP asymmetry signatures may not necessarily imply genuine leptonic CP violation originating from the PMNS matrix.
\\
Among the possible new-physics scenarios, non-unitary leptonic mixing provides a particularly attractive and phenomenologically relevant framework \cite{pavon2010unitarity}. Non-unitarity naturally arises in theories containing heavy neutral leptons, such as Type-I, inverse, and linear seesaw models, where the mixing between light and heavy neutrino states modifies the effective low-energy leptonic mixing matrix \cite{Antusch:2006vwa2}. In such scenarios, the effective mixing matrix governing neutrino oscillations is no longer exactly unitary and contains additional parameters and CP-violating phases \cite{martinez2020standard}. These new phases contribute directly to oscillation probabilities and may interfere with the standard oscillation amplitude, thereby producing fake CP-violating effects that mimic the signatures expected from a non-zero value of $\delta_{CP}$.
\\
An additional challenge arises from the interplay among non-unitarity effects, matter-induced asymmetries, and neutrino mass-ordering determination. Since all of these phenomena modify neutrino oscillation probabilities in an energy- and baseline-dependent manner, they can lead to a variety of parameter degeneracies. In particular, specific combinations of the standard CP phase, non-unitary phases, and mass ordering may generate nearly identical oscillation signatures. As a result, a normal-ordering solution with one set of non-unitary parameters may reproduce the event spectrum expected for an inverted-ordering solution with a different set of phases. Such hierarchy-CP-non-unitarity degeneracies can significantly reduce the sensitivity of individual experiments to both CP violation and mass ordering \cite{de2022theory}.
\\
To resolve these ambiguities, the complementarity between the Deep Underground Neutrino Experiment (DUNE) and Hyper-Kamiokande (Hyper-K) provides a particularly powerful strategy. Although both experiments primarily measure the $\nu_\mu \rightarrow \nu_e$ appearance channel, they operate in significantly different oscillation regimes owing to their distinct baselines and neutrino energy spectra. DUNE, with its long baseline of approximately $1300~\mathrm{km}$ and neutrino energies around the first oscillation maximum ($E\sim2$--$4~\mathrm{GeV}$), experiences substantial matter effects arising from coherent forward scattering of neutrinos in the Earth's crust. These matter-induced modifications strongly depend on the sign of $\Delta m_{31}^{2}$, thereby providing excellent sensitivity to the neutrino mass ordering while simultaneously enhancing the capability to distinguish genuine CP-violating effects from matter-induced asymmetries.
\\
In contrast, Hyper-Kamiokande operates at a much shorter baseline of approximately $295~\mathrm{km}$ with a neutrino beam peaked around $0.6~\mathrm{GeV}$, where matter effects are comparatively weak and the oscillation probabilities remain close to their vacuum behavior. Consequently, Hyper-K provides a precise measurement of the intrinsic CP-violating component of the oscillation probability with minimal contamination from matter effects. However, this also makes Hyper-K more susceptible to degeneracies between the standard Dirac phase $\delta_{CP}$ and the additional CP phases arising from non-unitary leptonic mixing.
\\
Because the two experiments probe different combinations of baseline, neutrino energy, and matter potential, the correlations among the oscillation parameters are markedly different in each case. Parameter combinations involving $\delta_{CP}$, the non-unitary parameters $(|\alpha_{21}|,\phi_{21})$, and the neutrino mass ordering that remain allowed by one experiment generally fail to reproduce the event spectrum observed by the other. A combined DUNE+Hyper-K analysis therefore provides significantly stronger constraints than either experiment alone by exploiting these complementary oscillation signatures. In particular, the joint analysis is expected to resolve a substantial fraction of the hierarchy--CP--non-unitarity degeneracies, thereby improving the determination of both the standard leptonic CP phase and the additional non-unitary parameters. This complementarity constitutes one of the principal motivations for performing a combined analysis in the present work.
\\
In this work, we investigate the impact of non-unitary neutrino mixing on CP violation and mass ordering measurements at DUNE and Hyper-Kamiokande. Particular attention is devoted to the non-unitary parameter $\alpha_{21}$ and its associated CP phase, which can generate significant interference effects in the $\nu_\mu \rightarrow \nu_e$ appearance channel. We evaluate the degradation of CP violation sensitivity, study the emergence of hierarchy-CP-non-unitarity degeneracies, and examine the capability of DUNE and Hyper-Kamiokande to distinguish genuine leptonic CP violation from fake CP asymmetries induced by non-unitary mixing. By performing a combined analysis of the two experiments, we identify regions of parameter space where the complementarity between different baselines and matter effects substantially improves the determination of both the leptonic CP phase and the neutrino mass ordering.

\section{Non-Unitary Framework}

In the standard three-flavor framework, neutrino oscillations arise because the flavor eigenstates produced and detected via weak interactions are coherent quantum superpositions of the neutrino mass eigenstates that propagate with different phases \cite{volpe2024neutrinos}. The flavor states $(\nu_e,\nu_\mu,\nu_\tau)$ are related to the mass eigenstates $(\nu_1,\nu_2,\nu_3)$ through the unitary Pontecorvo-Maki-Nakagawa-Sakata (PMNS) mixing matrix, $U_{\rm PMNS}$.
\begin{equation}
    |\nu_{\alpha}\rangle = \sum_{i=1}^{3} U_{\alpha i}^* |\nu_i\rangle, \hspace{1cm} \nu_{\alpha} = e, \mu, \tau      .
\end{equation}

Within this framework, neutrino oscillations are completely described by three mixing angles $(\theta_{12}, \theta_{13}, \theta_{23})$, two independent mass-squared differences $(\Delta m_{21}^{2}, \Delta m_{31}^{2})$, and a single Dirac CP-violating phase $\delta_{CP}$\cite{iizuka2015parametrization, nakamura2018neutrino}.
\\
Many well-motivated extensions of the Standard Model introduce heavy neutral leptons (HNLs) that mix with the three active neutrinos. Such heavy states naturally emerge in Type-I, inverse, and linear seesaw mechanisms, which provide an elegant explanation for the observed smallness of neutrino masses. In the complete theory containing both light and heavy neutrino states, the full leptonic mixing matrix remains unitary. However, after integrating out the heavy degrees of freedom, the effective mixing matrix describing the propagation of the light neutrinos is no longer exactly unitary \cite{PhysRevD.92.053009,Dutta:2019hmb,Soumya:2021dmy,Martinez-Soler:2018lcy,Dutta:2016czj,Tortola:2020plq,Blennow:2016jkn,Escrihuela:2016ube,fong2023theoretical}.

The low-energy effective leptonic mixing matrix can be parameterized as

\begin{equation}
N=(I-\alpha)U_{\rm PMNS},
\end{equation}

where $U_{\rm PMNS}$ denotes the standard unitary Pontecorvo-Maki-Nakagawa-Sakata (PMNS) matrix and $\alpha$ encodes deviations from exact unitarity arising from heavy-light neutrino mixing. The non-unitarity matrix can be written as

\begin{equation}
\alpha=
\begin{pmatrix}
\alpha_{ee} & \alpha_{e\mu} & \alpha_{e\tau} \\
\alpha_{\mu e}^{*} & \alpha_{\mu\mu} & \alpha_{\mu\tau} \\
\alpha_{\tau e}^{*} & \alpha_{\tau \mu}^{*} & \alpha_{\tau\tau}
\end{pmatrix},
\label{aplha-matrix}
\end{equation}
where the diagonal elements ($\alpha_{ee}$, $\alpha_{\mu\mu}$, $\alpha_{\tau\tau}$) modify the normalization of the neutrino flavor states, whereas the off-diagonal elements induce flavor-changing effects and introduce additional sources of CP violation beyond the standard PMNS framework.

In the present work, only the off-diagonal non-unitary parameter $\alpha_{\mu e}$ is considered to be non-zero, and it is parameterized as
\begin{equation}
\alpha_{\mu e}=|\alpha_{\mu e}|e^{i\phi_{\mu e}},
\label{alpha-phase}
\end{equation}

where $|\alpha_{\mu e}|$ denotes the magnitude of the non-unitary parameter and $\phi_{\mu e}$ represents the additional CP-violating phase associated with non-unitary leptonic mixing.\\
In this scenario the flavor eigenstates of neutrino are related to the neutrino mass eigenstates through

\begin{equation}
|\nu_{\alpha}\rangle = \sum_{i=1}^{3} N_{\alpha i}^* |\nu_i\rangle, \hspace{1cm} \nu_{\alpha} = e, \mu, \tau .
\end{equation}
In matter, neutrino propagation is governed by the effective Hamiltonian in the mass basis,
\begin{equation}
\label{eq:effectiveH}
H_{\rm mass}=\frac{1}{2E}
\begin{pmatrix}
0 & 0 & 0\\
0 & \Delta m_{21}^{2} & 0\\
0 & 0 & \Delta m_{31}^{2}
\end{pmatrix}
+
N^\dagger
\begin{pmatrix}
V_{CC}+V_{NC} & 0 & 0\\
0 & V_{NC} & 0\\
0 & 0 & V_{NC}
\end{pmatrix}
N
\end{equation}

where

\begin{equation}
V_{CC}=\sqrt{2}G_F N_e
\end{equation}
and
\begin{equation}
V_{NC}=
-\frac{\sqrt{2}}{2}G_F N_n
\end{equation}
are the charged-current and neutral-current matter potentials, respectively.
The evolution operator is given by
\begin{equation}
S_{\rm mass}(L)=\exp\left(-iH_{\rm mass}L\right).
\end{equation}

The corresponding flavor-basis evolution matrix is

\begin{equation}
S_{\rm flavor}(L)=NS_{\rm mass}(L)N^\dagger .
\end{equation}

The oscillation probability is therefore

\begin{equation}
P_{\alpha\beta}^{\rm NU}=\frac{
\left|
\left[
S_{\rm flavor}(L)
\right]_{{\beta\alpha}}
\right|^2
}
{
(NN^\dagger)_{\alpha\alpha}
(NN^\dagger)_{\beta\beta}
}.
\end{equation}

For the appearance channel relevant to DUNE and Hyper-Kamiokande,

\begin{equation}
P_{\mu e}^{\rm NU}=\frac{
\left|
\left[
N
e^{-iH_{\rm mass}L}
N^\dagger
\right]_{e\mu}
\right|^2
}
{
(NN^\dagger)_{\mu\mu}
(NN^\dagger)_{ee}
}
\end{equation}
The additional CP-violating phase associated with the non-unitary parameter, defined in Equation.~(\ref{alpha-phase}), enters the effective leptonic mixing matrix and consequently modifies the neutrino oscillation probabilities. As a result, observable CP-violating effects may arise even when the standard Dirac CP phase assumes the CP-conserving values, $\delta_{CP}=0$ or $\pi$.
Consequently, an experimentally observed CP asymmetry does not necessarily imply genuine leptonic CP violation originating from the PMNS matrix. \\
The appearance channel ${\nu_\mu \rightarrow \nu_e}$, which constitutes the primary probe of CP violation in DUNE and Hyper-Kamiokande, is particularly sensitive to the parameter ${\alpha_{21}}$. The corresponding phase ${\phi_{21}}$ may interfere with the standard oscillation amplitude and produce degeneracies between genuine and non-unitary CP effects. Such degeneracies can significantly bias the extraction of ${\delta_{CP}}$ and reduce the sensitivity to neutrino mass ordering.
\\
Recent studies have emphasized that future long-baseline experiments may enter a precision regime in which non-unitarity effects become comparable to statistical uncertainties. In this context, it becomes essential to distinguish between three different sources of CP asymmetry: genuine CP violation associated with the PMNS phase, fake CP violation induced by non-unitary phases, and matter-induced asymmetries arising from neutrino propagation through the Earth.
\\
An additional and largely unexplored aspect concerns the correlation between non-unitarity and mass-ordering determination. Since both matter effects and non-unitary corrections modify the oscillation probability in a baseline-dependent manner, specific combinations of ${\alpha_{21}}$, ${\alpha_{31}}$, and their associated phases can mimic the oscillation pattern of the opposite mass ordering. This opens the possibility that future measurements could simultaneously suffer from hierarchy–CP and non-unitarity–CP degeneracies.
\\
In the present work, we investigate these effects within a unified framework and explore whether combined measurements from DUNE and Hyper-Kamiokande can separate genuine CP violation from new-physics-induced fake CP signals. Particular emphasis is placed on identifying parameter regions where the observed CP asymmetry is dominated by non-unitarity effects and on constructing observables capable of discriminating between the two scenarios.

\section{Experimental Setup}

The present analysis is performed using the Global Long Baseline Experiment Simulator (GLoBES) \cite{Huber:2004ka,Huber:2007ji} with the official configuration files of the Deep Underground Neutrino Experiment (DUNE) \cite{abi2021experiment} and Hyper-Kamiokande (Hyper-K) \cite{huber2002superbeams,itow2001jhf,ishitsuka2005resolving,paschos2002neutrino,messier1999evidence} . The simulations incorporate the detector response, neutrino fluxes, interaction cross sections, migration matrices, detector efficiencies, and systematic uncertainties as specified in the respective Technical Design Reports (TDRs). To study the impact of non-unitary leptonic mixing, the standard GLoBES probability engine is replaced by a modified probability engine implementing the non-unitary oscillation formalism.
\\
For DUNE, a baseline of approximately $1300~\mathrm{km}$ and a $40~\mathrm{kt}$ liquid argon time projection chamber (LArTPC) far detector are considered. The experiment is simulated with a total exposure of 10 years, corresponding to 5 years in neutrino mode and 5 years in antineutrino mode. Hyper-Kamiokande is simulated with a baseline of $295~\mathrm{km}$ using a $187~\mathrm{kt}$ fiducial water Cherenkov detector and a total exposure of 10 years, divided into 2.5 years in neutrino mode and 7.5 years in antineutrino mode, following the design configuration adopted in the Hyper-K Technical Design Report.
\\
Unless otherwise stated, the oscillation parameters are fixed to their current global best-fit values from NuFit~6.0 \cite{Esteban:2024eli}. The analysis assumes Normal Ordering (NO) as the true mass ordering, while both Normal and Inverted Ordering (IO) are considered in the fit. 
The sensitivity is evaluated through a standard chi-square analysis.

The test statistic is defined as

\begin{equation}
\chi^2
=
\sum_i
\frac{
\left(
N_i^{obs}
-
N_i^{th}
\right)^2
}
{\sigma_i^2}
+
\chi^2_{prior}.
\end{equation}

Marginalization is performed over oscillation parameters, systematic uncertainties, and non-unitary parameters.

\section{Genuine and Fake CP Violation}
\label{sec:genuine_fake_cp}
A key objective of future oscillation experiments is the measurement of leptonic CP violation through the comparison of neutrino and antineutrino oscillation probabilities. The CP asymmetry is defined as
\begin{equation}
A_{CP} = P(\nu_\mu\rightarrow\nu_e) - P(\bar{\nu}_\mu\rightarrow\bar{\nu}_e).
\end{equation}
In the standard three-flavor framework, this asymmetry originates from the Dirac CP phase $\delta_{CP}$ contained in the PMNS matrix. In the presence of non-unitary mixing, however, additional CP-violating phases contribute to the oscillation probability, and the total asymmetry may be schematically decomposed as
\begin{equation}
A_{CP} = A_{CP}^{\rm PMNS} + A_{CP}^{\rm NU} + A_{CP}^{\rm matter},
\end{equation}
where $A_{CP}^{\rm PMNS}$ represents genuine leptonic CP violation generated by the standard phase $\delta_{CP}$, $A_{CP}^{\rm NU}$ arises from the additional phases associated with non-unitarity, and $A_{CP}^{\rm matter}$ denotes the asymmetry induced by neutrino propagation through matter~\cite{Bernabeu2018DisentanglingGF, Escrihuela:2016ube, fernandez2007cp}. An important consequence of non-unitary mixing is that
\begin{equation}
A_{CP}^{\rm NU} \neq 0 \qquad \text{for} \qquad \delta_{CP} = 0 
\;\text{or}\; \pi,
\end{equation}
provided that one or more non-unitary phases are nonzero. An experimentally observed CP asymmetry therefore does not by itself imply genuine PMNS-induced CP violation : the measured signal may instead contain contributions from new CP phases associated with heavy neutral leptons, producing what is commonly termed \emph{fake CP violation}.

\begin{figure}[htbp]
    \centering
    \begin{subfigure}[t]{0.49\textwidth}
        \centering
        \includegraphics[width=\linewidth]{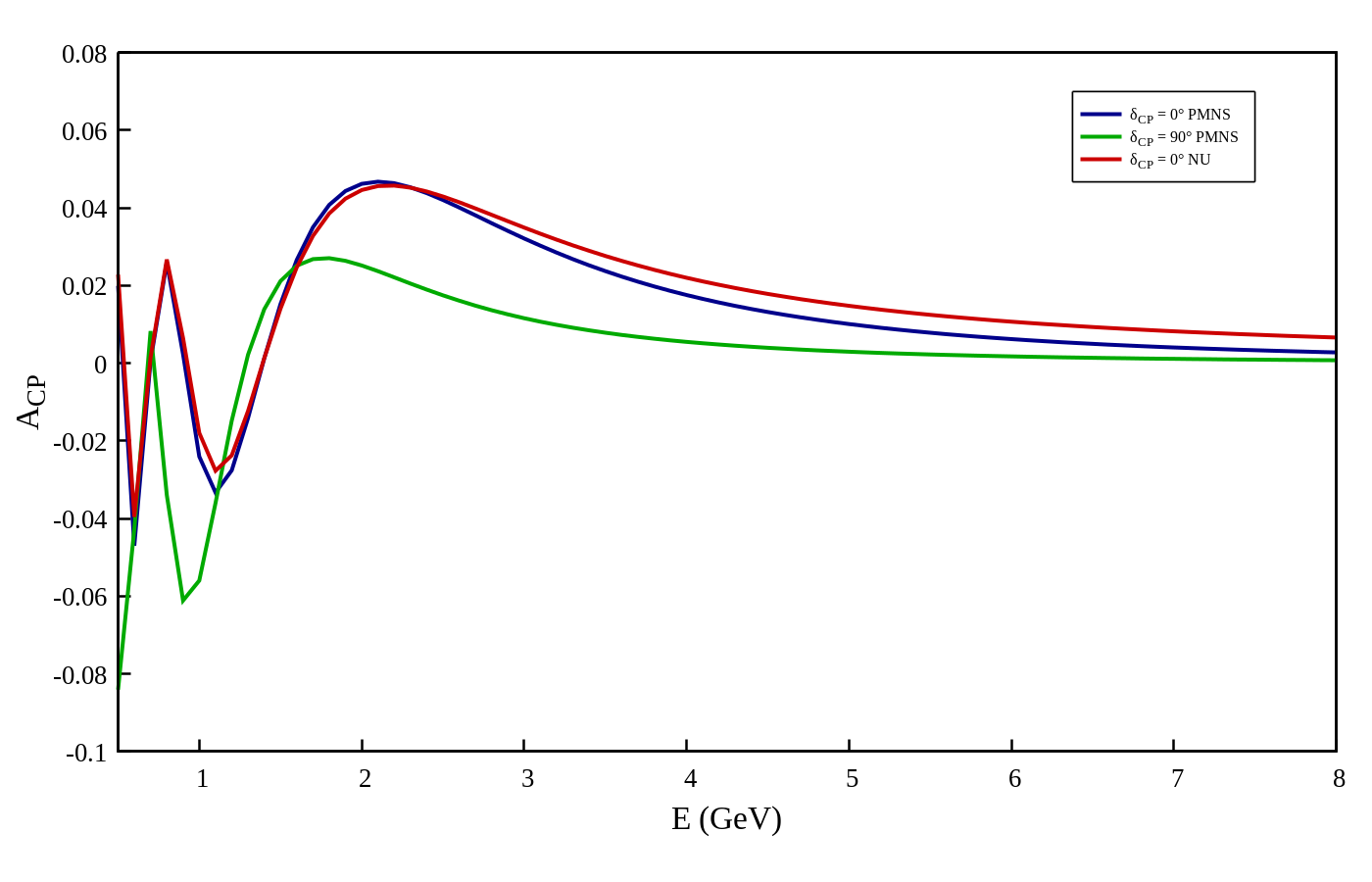}
        \caption{DUNE ($L = 1300$~km): $\delta_{CP} = 0^\circ$ PMNS (blue), 
        $\delta_{CP} = 90^\circ$ PMNS (green), and non-unitary benchmark 
        $\alpha_{21}=0.01$, $\varphi_{21}=-90^\circ$ at $\delta_{CP}=0^\circ$ 
        (red).}
        \label{fig:ACP_DUNE}
    \end{subfigure}
    \hfill
    \begin{subfigure}[t]{0.49\textwidth}
        \centering
        \includegraphics[width=\linewidth]{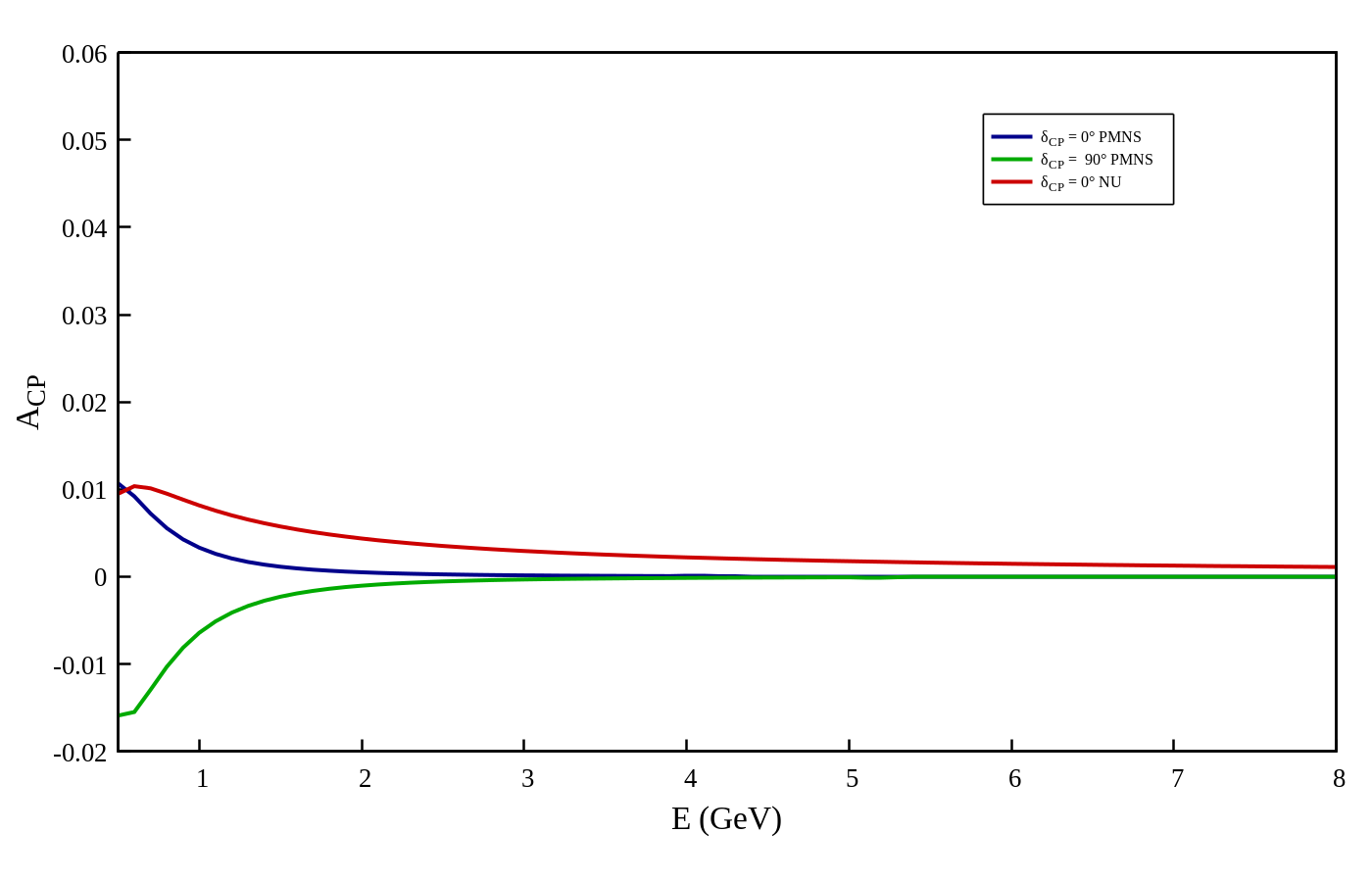}
        \caption{Hyper-Kamiokande ($L = 295$~km): same color coding as 
        panel~(a).}
        \label{fig:ACP_T2HK}
    \end{subfigure}
    \caption{Energy dependence of the CP asymmetry $A_{CP}(E) = P(\nu_\mu\rightarrow\nu_e) - P(\bar{\nu}_\mu\rightarrow\bar{\nu}_e)$ for DUNE and Hyper-Kamiokande, comparing the standard PMNS prediction at two representative values of $\delta_{CP}$ against the non-unitary benchmark at $\alpha_{21}=0.01$, $\varphi_{21}=-90^\circ$.}
    \label{fig:Probability_Asymmetry}
\end{figure}

\subsection{Energy dependence of the CP asymmetry}

Figures~\ref{fig:ACP_DUNE} and \ref{fig:ACP_T2HK} compare $A_{CP}(E)$ for DUNE ($L=1300$~km) and Hyper-Kamiokande ($L=295$~km) in the standard PMNS and non-unitary scenarios. The two experiments exhibit qualitatively different behavior reflecting their very different baselines.
\\
For DUNE, the long baseline of $L=1300$~km produces substantial matter effects that generate a rich low-energy structure in $A_{CP}(E)$. Below $E\approx1.5$~GeV, the charged-current matter potential $V_{cc}$ dominates the oscillation Hamiltonian, and $A_{CP}^{\rm matter}$ becomes the largest contribution to the total asymmetry. In this region the $\delta_{CP}=0^\circ$ PMNS and non-unitary curves are nearly indistinguishable: since both are governed by the same matter-induced term, sensitivity to both the genuine PMNS phase and the non-unitary phase $\varphi_{21}$ is simultaneously suppressed. As the neutrino energy increases toward the first oscillation maximum near $E\approx2.0$~GeV, where the vacuum oscillation phase $\Delta m^2_{31}L/4E=\pi/2$ and $P(\nu_\mu\rightarrow\nu_e)$ is maximized; the matter-induced contribution diminishes relative to the intrinsic CP-violating terms, and the three curves begin to separate visibly. The 
green ($\delta_{CP}=90^\circ$ PMNS) curve falls below the blue ($\delta_{CP}=0^\circ$ PMNS) curve, while the red (non-unitary) curve tracks closely above blue. At this oscillation maximum, where DUNE's $\delta_{CP}$ sensitivity is strongest, the $\delta_{CP}=0^\circ$ PMNS asymmetry peaks at $0.04674$ near $E\approx2.1$~GeV, and the non-unitary curve peaks at $A_{CP}^{\rm NU}(\delta_{CP}=0^\circ)\approx0.04573$ near $E\approx2.2$~GeV, only $0.00101$ lower in absolute terms demonstrating that the non-unitary phase $\varphi_{21}=-90^\circ$ generates a CP asymmetry of nearly identical magnitude to the CP-conserving PMNS baseline, even in the absence of any genuine Dirac CP violation.
\\
For Hyper-Kamiokande, the shorter baseline results in negligible matter effects throughout the energy range, and $A_{CP}$ is sensitive primarily to the intrinsic and non-unitary CP phases. The curves show a simpler structure without the low-energy oscillatory features present at DUNE; the $\delta_{CP}=0^\circ$ PMNS and non-unitary curves track closely together across the full energy range, while $\delta_{CP}=90^\circ$ PMNS is separated below them. The non-unitary peak of $0.01036$ near $E\approx0.6$~GeV is nearly identical to the $\delta_{CP}=0^\circ$ PMNS peak of $0.01068$ at $E\approx0.5$~GeV. This near-degeneracy reflects the small magnitude of the non-unitary correction at $\alpha_{21}=0.01$; the fake contribution is a perturbation of order $|\alpha_{21}|$ on the 
PMNS baseline, insufficient to produce a visible separation in the direct amplitude of $A_{CP}$ at Hyper-Kamiokande's peak-sensitivity energy, as confirmed quantitatively by the fake-to-genuine ratio of $4.5\%$ in Table~\ref{tab:acp}.

\subsection{Quantifying genuine and fake contributions}
A more precise picture of the contamination emerges from comparing the genuine CP asymmetry swing with the fake non-unitary excess at the same energy. The genuine swing quantifies the change in the CP asymmetry, $A_{\mathrm{CP}}$, when genuine CP violation is varied from the CP-conserving case ($\delta_{\mathrm{CP}}=0^\circ$) to maximal CP violation ($\delta_{\mathrm{CP}}=90^\circ$). In contrast, the fake excess quantifies the change in $A_{\mathrm{CP}}$ when non-unitary mixing is introduced while keeping $\delta_{\mathrm{CP}}=0^\circ$, ensuring that the resulting asymmetry arises solely from the non-unitary phase rather than from the standard Dirac CP phase, $\delta_{\mathrm{CP}}$.
\\
We define the genuine swing as 
$|A_{CP}(\delta_{CP}=0^\circ) - A_{CP}(\delta_{CP}=90^\circ)|$ and the fake excess as 
$|A_{CP}^{\rm NU}(\delta_{CP}=0^\circ) - A_{CP}^{\rm PMNS}(\delta_{CP}=0^\circ)|$, both evaluated from the raw probability files. Table~\ref{tab:acp} summarizes these quantities at the energy where the genuine swing is maximal and at the energy window where the genuine signal is locally suppressed.

\begin{table}[htbp]
\centering
\caption{Genuine CP asymmetry swing, fake non-unitary excess, and their ratio at representative neutrino energies for DUNE and Hyper-Kamiokande, evaluated at the benchmark non-unitary parameters
$\alpha_{21}=0.01$, $\varphi_{21}=-90^\circ$.
The genuine swing is $|A_{CP}(\delta_{CP}=0^\circ)-A_{CP}(\delta_{CP}=90^\circ)|$
and the fake excess is
$|A_{CP}^{\rm NU}(\delta_{CP}=0^\circ)-A_{CP}^{\rm PMNS}(\delta_{CP}=0^\circ)|$,
both computed from the raw oscillation probability files. A ratio exceeding unity indicates that the fake contribution alone exceeds the genuine CP-violating signal at that energy.}
\label{tab:acp}
\renewcommand{\arraystretch}{1.4}
\begin{tabular}{llccc}
\hline\hline
Experiment & Energy regime & Genuine swing & Fake excess & Ratio \\
\hline
\multirow{2}{*}{DUNE}
 & Oscillation maximum ($2.4$~GeV)      
   & $0.02485$ & $0.00041$ & $1.7\%$ \\
 & Suppressed-genuine window ($1.6$~GeV) 
   & $0.00148$ & $0.00191$ & $129\%$ \\
\hline
\multirow{3}{*}{Hyper-Kamiokande}
 & Oscillation maximum ($0.5$~GeV)       
   & $0.02658$ & $0.00119$ & $4.5\%$ \\
 & First suppressed window ($1.5$~GeV)   
   & $0.00343$ & $0.00460$ & $134\%$ \\
 & Second suppressed window ($2.0$~GeV)  
   & $0.00300$ & $0.00387$ & $129\%$ \\
\hline\hline
\end{tabular}
\end{table}

At the energy where each experiment's genuine swing is maximal, the fake non-unitary contribution is small: $1.7\%$ of the genuine signal at DUNE and $4.5\%$ at Hyper-Kamiokande. This confirms that, at the energies where these experiments are designed to be most sensitive to $\delta_{CP}$, non-unitary contamination at the level of $\alpha_{21}=0.01$ is a minor perturbation rather than a serious systematic effect.
\\
However, this favorable comparison does not hold across the full energy spectrum. Near $E \approx 1.5$--$1.6$~GeV at both experiments, the genuine PMNS asymmetry passes through a local suppression where the $\delta_{CP}=0^\circ$ and $\delta_{CP}=90^\circ$ curves happen to be 
close to each other, reducing the genuine swing to $0.00148$ at DUNE and $0.00343$ at Hyper-Kamiokande. The fake non-unitary contribution does not undergo a corresponding suppression at these energies, remaining at $0.00191$ and $0.00460$ respectively. As a result, the fake-to-genuine ratio exceeds unity in this window, reaching $129\%$ at DUNE and $134\%$ 
at Hyper-Kamiokande. A second such window exists for Hyper-Kamiokande near $E \approx 2.0$~GeV, where the genuine swing falls to $0.00300$ while the fake excess remains $0.00387$ ($129\%$). These energy windows represent the specific regions of the spectrum where an $A_{CP}$-based 
CP violation measurement is most vulnerable to contamination from non-unitary effects: an observed asymmetry in these bins could be dominated by fake rather than genuine CP violation, and could not be distinguished from a genuine PMNS signal on the basis of $A_{CP}$ alone.

\subsection{Phase-sign dependence and visual in-distinguishability}

\begin{figure}[htbp]
    \centering
    \includegraphics[width=0.85\linewidth]{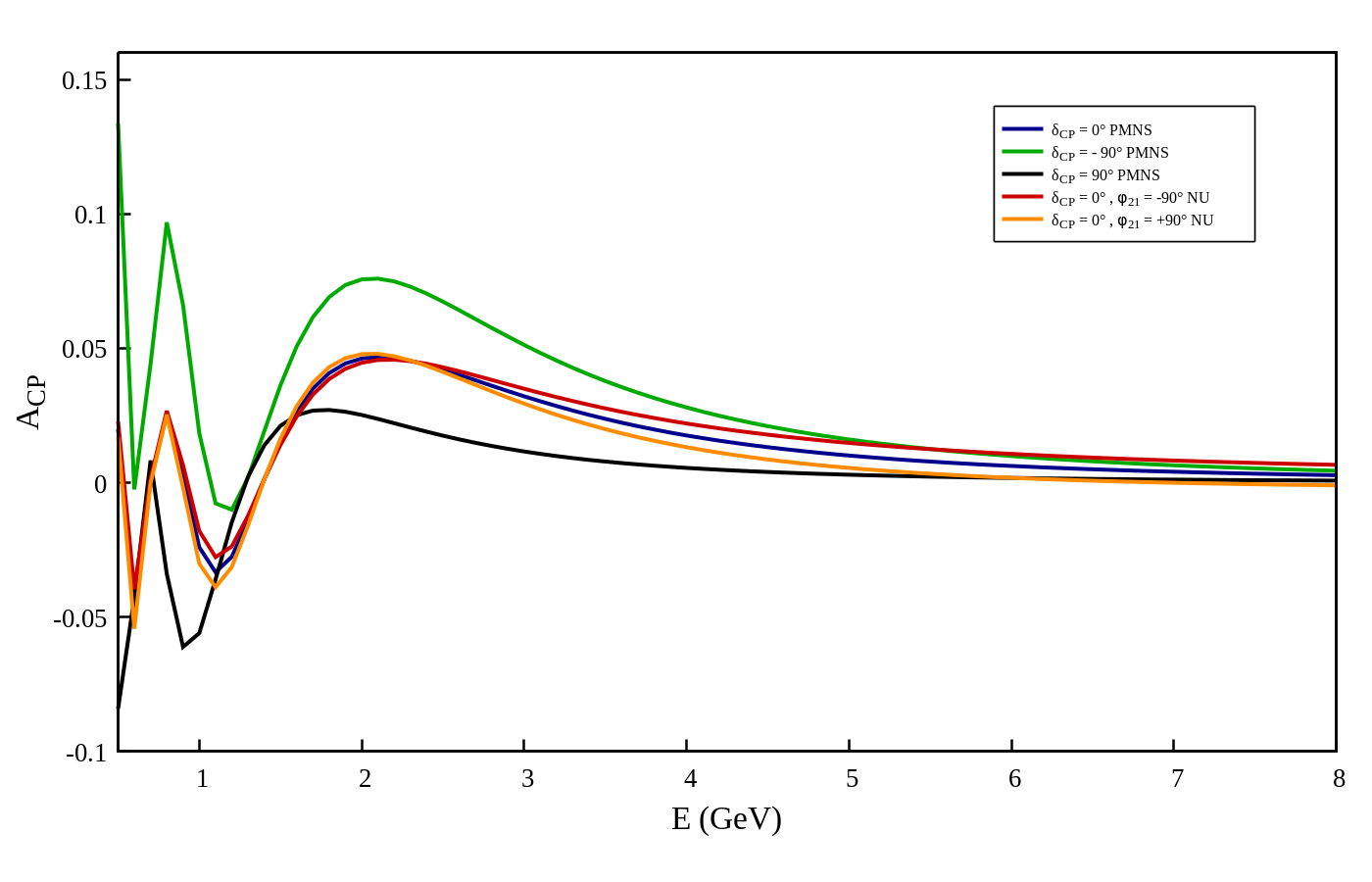}
    \caption{$A_{CP}(E)$ at DUNE for $\delta_{CP} = 0^\circ$ (blue), $\delta_{CP} = -90^\circ$ (green), and $\delta_{CP} = +90^\circ$ (black) PMNS, together with the non-unitary benchmark at $\varphi_{21} = -90^\circ$ (red) and $\varphi_{21} = +90^\circ$ (orange), both at $\delta_{CP}=0^\circ$. Neither non-unitary curve is visually distinguishable from the $\delta_{CP}=0^\circ$ PMNS curve in the broad-peak region.}
    \label{fig:ACP_signcheck}
\end{figure}

Figure~\ref{fig:ACP_signcheck} extends this comparison by plotting both signs of the non-unitary phase ($\varphi_{21} = \pm90^\circ$) alongside all three PMNS benchmarks ($\delta_{CP} = 0^\circ, \pm90^\circ$). Throughout the dominant broad-peak region ($E \approx 1.7$--$4$~GeV), both non-unitary curves (red and orange) remain visually indistinguishable from the $\delta_{CP}=0^\circ$ PMNS curve (blue), regardless of the sign of $\varphi_{21}$. By contrast, the genuine CP-violating benchmarks $\delta_{CP}=-90^\circ$ (green) and $\delta_{CP}=+90^\circ$ (black) are clearly and symmetrically separated from this cluster on opposite sides, as expected from the CP-odd nature of the $\delta_{CP}$ dependence.
\\
This result has two important implications-- first, it confirms that the visual near-degeneracy between PMNS and non-unitary predictions is not an accident of a particular phase convention : the two signs of $\varphi_{21}$ produce indistinguishable $A_{CP}$ curves in the broad-peak region, so neither the sign nor the presence of this non-unitary benchmark can be inferred from the shape of $A_{CP}$ in this energy window. Second, it demonstrates that $A_{CP}$ is an effective discriminator of genuine $\delta_{CP}$; the separation between the $\delta_{CP}=0^\circ$ and $\delta_{CP}=\pm90^\circ$ PMNS curves is large and clearly measurable, while the non unitary shift from the $\delta_{CP}=0^\circ$ baseline is invisible at this scale. The reason is straightforward as the absolute fake contribution at this benchmark ($\alpha_{21}=0.01$) is small and approximately constant across energy, so it is simply swamped by the large genuine $\delta_{CP}$ signal in the broad-peak region. It becomes visible only where the genuine signal is locally suppressed, precisely the energy windows identified in Table~\ref{tab:acp}.

\subsection{Joint-experiment asymmetry difference}

\begin{figure*}[t]
    \centering
    \includegraphics[width=0.75\textwidth]{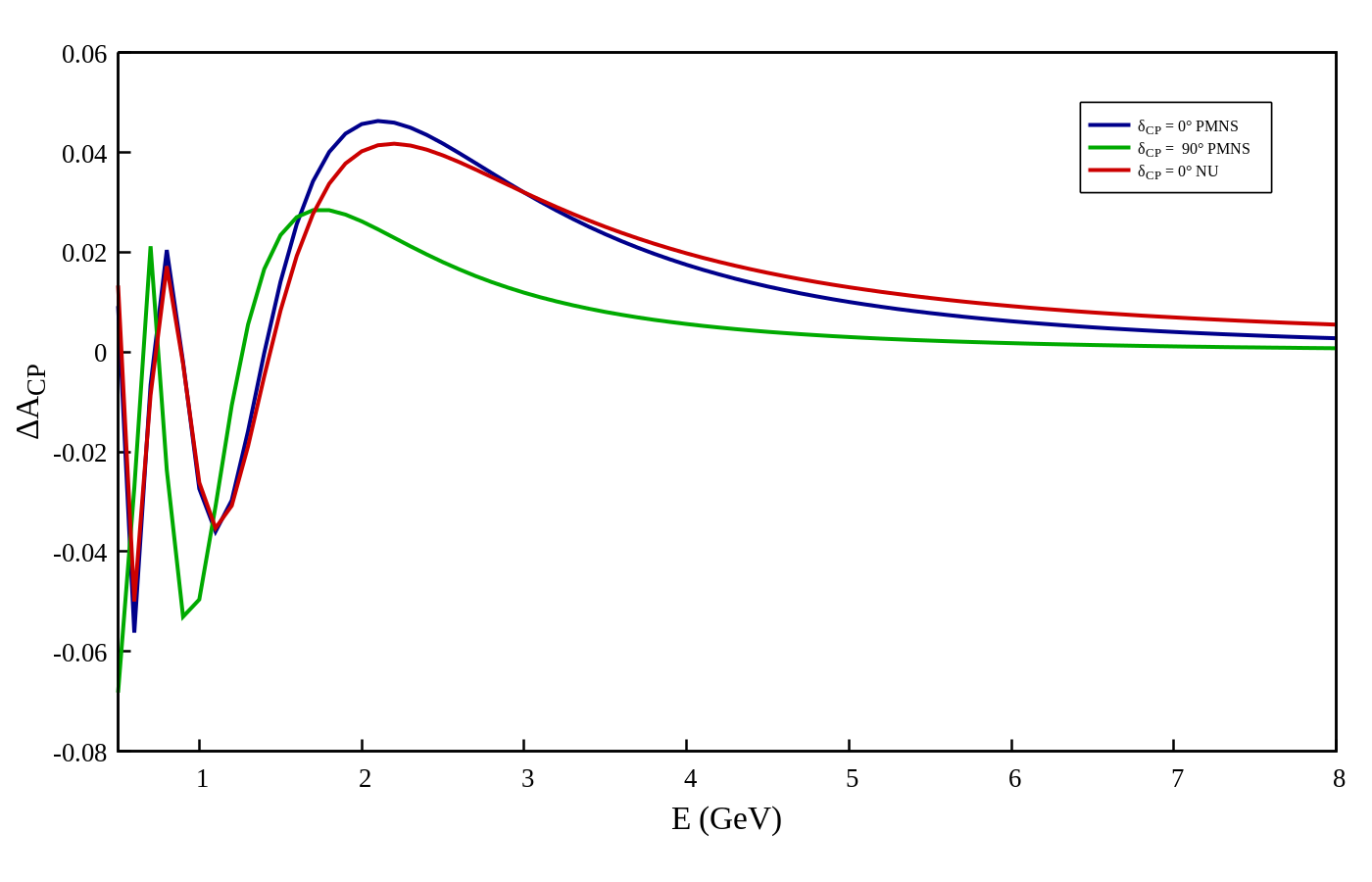}
    \caption{The joint CP asymmetry difference $\Delta A_{CP}(E) = A_{CP}^{\rm DUNE}(E) - A_{CP}^{\rm HK}(E)$ for $\delta_{CP}=0^\circ$ PMNS (blue), $\delta_{CP}=90^\circ$ PMNS (green), and the non-unitary benchmark at $\alpha_{21}=0.01$, $\varphi_{21}=-90^\circ$ (red). The non-unitary curve tracks the $\delta_{CP}=0^\circ$ PMNS curve closely throughout the broad-peak region, confirming that this observable does not add discriminating power against non-unitarity beyond what is available from $A_{CP}$ alone.}
    \label{fig:Delta_A}
\end{figure*}

To test whether combining the two experiments provides additional discriminating power, we define the joint observable
\begin{equation}
\Delta A_{CP}(E) = A_{CP}^{\rm DUNE}(E) - A_{CP}^{\rm HK}(E).
\end{equation}
The complementary baselines of DUNE and Hyper-Kamiokande enter $\Delta A_{CP}$ through their different matter effects and $L/E$ regimes, so one might hope that the non-unitary and PMNS predictions separate more clearly in this difference than in either individual asymmetry.
\\
Figure~\ref{fig:Delta_A} shows that this is not the case in the broad-peak region. The non-unitary curve peaks at $+0.0417$ near $E \approx 2.2$~GeV, close to the $\delta_{CP}=0^\circ$ PMNS peak of $+0.0463$ at $E \approx 2.1$~GeV, differing by only $\sim10\%$ between them. Both curves are well separated from the $\delta_{CP}=90^\circ$ PMNS peak of $+0.0284$ at $E \approx 1.7$~GeV ($\sim60\%$ difference). The non-unitary and $\delta_{CP}=0^\circ$ PMNS curves also share qualitatively similar shapes throughout the full energy range shown, including the low-energy oscillatory feature below $1.5$~GeV that originates from DUNE's matter-resonance structure. Consequently, $\Delta A_{CP}$ does not provide additional discrimination against the non-unitary benchmark beyond what is already available from $A_{CP}$ alone; in the energy region where both signals are large, the fake contribution remains indistinguishable from the CP-conserving PMNS prediction, for the same reason as found in the individual asymmetries.
\\
\\
The results of this section establish a consistent and physically transparent picture. The CP asymmetry $A_{CP}$ is a powerful observable for measuring genuine $\delta_{CP}$; the separation between CP-conserving ($\delta_{CP}=0^\circ$) and maximally CP-violating ($\delta_{CP}=\pm90^\circ$) predictions is large and clearly visible at both DUNE and Hyper-Kamiokande throughout the oscillation-maximum region. Non-unitary mixing at the benchmark level of $\alpha_{21}=0.01$ produces a fake CP asymmetry that is small in absolute magnitude and indistinguishable from the $\delta_{CP}=0^\circ$ PMNS prediction in this region, irrespective of the sign of $\varphi_{21}$ and irrespective of whether the experiments are considered individually or in combination through $\Delta A_{CP}$.
\\
The contamination becomes significant, however, in specific energy windows where the genuine PMNS asymmetry is locally suppressed. At $E \approx 1.5$--$1.6$~GeV the fake-to-genuine ratio exceeds unity at both experiments ($129\%$ at DUNE, $134\%$ at Hyper-Kamiokande); a second such window exists for Hyper-Kamiokande near $E \approx 2.0$~GeV ($129\%$). In these specific bins, a measured CP asymmetry could be dominated by fake rather than genuine CP violation, and a standard $A_{CP}$ based analysis would be unable to distinguish the two on the basis of the asymmetry value alone. This motivates either constraining $\alpha_{21}$ independently via near-detector data before interpreting any asymmetry signal in these bins, or applying an energy-binned $\chi^2$ analysis that weights or excludes the identified contaminated regions; a strategy examined quantitatively in the following section.


\section{Parameter Degeneracies}

One of the most significant challenges in measuring leptonic CP violation in the presence of non-unitary mixing is the emergence of parameter degeneracies between the standard Dirac phase $\delta_{CP}$ and the non-unitary phase $\varphi_{21}$. These degeneracies arise because both phases contribute to the appearance probability $P(\nu_\mu\rightarrow\nu_e)$ through interference terms that can produce identical oscillation signatures at a given baseline and energy, making 
it impossible to attribute an observed CP asymmetry uniquely to either 
source without additional experimental information.
\\
A characteristic instance of this degeneracy can be written as
\begin{equation}
P_{\rm NU}(\delta_{CP}=0,\;\varphi_{21}\neq 0,\;V_{cc})
=
P_{\rm 3\nu}(\delta_{CP}\neq 0,\;V_{cc}),
\label{eq:degeneracy}
\end{equation}
where $V_{cc}=\sqrt{2}G_F N_e$ is the charged-current matter potential. Equation~\eqref{eq:degeneracy} states that a non-unitary scenario with no genuine CP violation ($\delta_{CP}=0$) but a non-zero phase $\varphi_{21}$ can reproduce the same oscillation probability as a standard three-flavor scenario with genuine CP violation ($\delta_{CP}\neq 0$) and no non-unitarity. As demonstrated in Sec.~\ref{sec:genuine_fake_cp}, this degeneracy is energy-dependent and most severe in specific energy windows where the genuine CP asymmetry is locally suppressed.
\\
The degeneracy is further compounded by the neutrino mass ordering. Matter effects at long baselines modify the oscillation probability differently for NO and IO, and these modifications depend on both $\delta_{CP}$ and $\varphi_{21}$. A three-way hierarchy--CP--non-unitarity degeneracy therefore arises when
\begin{equation}
P_{\mu e}^{\rm NO}(\delta_{CP},\;\varphi_{21})
\;\simeq\;
P_{\mu e}^{\rm IO}(\delta_{CP}^{\prime},\;\varphi_{21}^{\prime}),
\label{eq:hcn}
\end{equation}
where the primed parameters correspond to the wrong mass ordering with a different combination of CP phases. This is the most challenging degeneracy scenario : an experiment assuming the correct mass ordering may reconstruct the wrong CP phases, and one fitting the wrong ordering may absorb the error into a fake combination of $\delta_{CP}^{\prime}$ and $\varphi_{21}^{\prime}$.

\subsection*{Analysis method}

To map these degeneracies quantitatively, we compute the hierarchy--CP--non-unitarity $\chi^2$ surface over the full $(\delta_{CP}^{\rm test}/\pi,\;\varphi_{21}^{\rm test}/\pi)\in[-1,1]^2$ parameter space. The true event spectra are generated with NO and the following benchmark parameters are used \cite{Esteban:2024eli}:

To map these degeneracies quantitatively, we compute the hierarchy--CP--non-unitarity $\chi^2$ surface over the full $(\delta_{CP}^{\rm test}/\pi,\;\varphi_{21}^{\rm test}/\pi)\in[-1,1]^2$ parameter space. The true event spectra are generated assuming NO and the oscillation parameters are fixed to the current global best-fit values reported in Ref.~\cite{Esteban:2024eli}:
\begin{equation}
\begin{aligned}
\sin^2\theta_{12} &= 0.307,\quad
\sin^2\theta_{13} = 0.02195,\quad
\sin^2\theta_{23} = 0.561,\\
\delta_{CP}^{\rm true}/\pi &= -0.5,\quad
\Delta m^2_{21} = 7.49\times10^{-5}~{\rm eV}^2,\quad
\Delta m^2_{31} = 2.534\times10^{-3}~{\rm eV}^2,\\
|\alpha_{21}|^{\rm true} &= 0.01,\quad
\varphi_{21}^{\rm true} = 0.
\end{aligned}
\end{equation}
For each point $(\delta_{CP}^{test},\varphi_{21}^{ test})$ on a uniform grid with step size $0.1$~rad in both directions, the test hypothesis assumes the wrong mass ordering (IO, implemented by $(\Delta m^2_{31})^{\text{test}}<0$) and the $\chi^2$ is computed using \texttt{glbChiSys} for DUNE and Hyper-Kamiokande separately. The combined $\chi^2$ is formed as their sum:
\begin{equation}
\chi^2_{\rm combined} = \chi^2_{\rm DUNE} + \chi^2_{\rm HK}.
\end{equation}
At each grid point, an explicit brute-force marginalization is performed over three nuisance parameters : 
$|\alpha_{21}|^{\rm test}\in[0,\,0.02]$ (step $0.002$), 
$\sin^2\theta_{23}^{\rm test}\in[0.40,\,0.62]$ (step $0.02$), and 
$|\Delta m^2_{31}|^{\rm test}\in[2.40,\,2.60]\times10^{-3}$~eV$^2$ 
(step $2\times10^{-5}$~eV$^2$), taking the minimum $\chi^2$ over 
all combinations. The confidence contours are defined by
\begin{equation}
\Delta\chi^2 = \chi^2(\delta_{CP}^{\rm test},\varphi_{21}^{\rm test}) 
- \chi^2_{\rm min},
\end{equation}
with thresholds $\Delta\chi^2 = 2.30$, $6.18$, $11.83$ for the 
$1\sigma$, $2\sigma$, $3\sigma$ levels at two degrees of freedom.


\begin{figure*}[t]
\centering
\includegraphics[width=\textwidth]{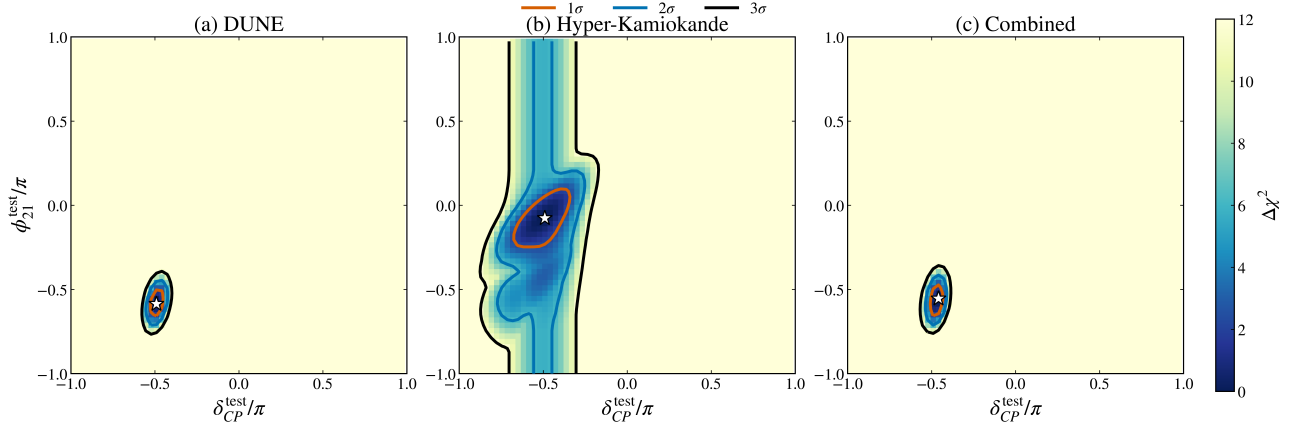}
\caption{Allowed regions in the 
$(\delta_{CP}^{\rm test}/\pi,\;\varphi_{21}^{\rm test}/\pi)$ parameter space from the hierarchy--CP--non-unitarity degeneracy analysis for (a)~DUNE, (b)~Hyper-Kamiokande, and (c)~combined DUNE + Hyper-Kamiokande. Contours show the $1\sigma$ (orange), $2\sigma$ (blue), and $3\sigma$ (black) confidence levels for two degrees of freedom ($\Delta\chi^2 = 2.30$, $6.18$, $11.83$). The color scale shows $\Delta\chi^2$ at each grid point. The white star marks the global best-fit point in each panel. True parameters: NO, $\delta_{CP}^{\rm true}/\pi=-0.5$, $|\alpha_{21}|^{\rm true}=0.01$, $\varphi_{21}^{\rm true}=0$. Test hypothesis: IO, marginalized over $|\alpha_{21}|$, $\sin^2\theta_{23}$, $|\Delta m^2_{31}|$.}
\label{fig:Degeneracy}
\end{figure*}

The global $\chi^2$ minima obtained from the scan are

\begin{equation}
(\chi^2)_{\rm min}^{\rm DUNE} = 485.853,\qquad
(\chi^2)_{\rm min}^{\rm HK}   = 4.404,  \qquad
(\chi^2)_{\rm min}^{\rm combined} = 507.867,
\end{equation}
with best-fit points at
\begin{equation}
\begin{aligned}
{\rm DUNE:}\quad
&\delta_{CP}^{\rm test}/\pi = -0.4907,\quad
\varphi_{21}^{\rm test}/\pi = -0.5862,\\
{\rm HK:}\quad
&\delta_{CP}^{\rm test}/\pi = -0.4907,\quad
\varphi_{21}^{\rm test}/\pi = -0.0769,\\
{\rm Combined:}\quad
&\delta_{CP}^{\rm test}/\pi = -0.4589,\quad
\varphi_{21}^{\rm test}/\pi = -0.5544.
\end{aligned}
\end{equation}
All three best-fit $\delta_{CP}$ values lie within $5\%$ of the true value $\delta_{CP}^{\rm true}/\pi=-0.5$, confirming that the genuine CP phase is recovered correctly even in the presence of the non-unitarity degeneracy. The best-fit $\varphi_{21}$ values, however, differ significantly from the true value $\varphi_{21}^{\rm true}=0$, reflecting the residual freedom in the non-unitary phase that each experiment cannot resolve independently when the wrong mass ordering is assumed.
\\
\\
\textbf{DUNE alone} (panel a). The $1\sigma$, $2\sigma$, and $3\sigma$ 
allowed regions are compact ellipses confined to
\begin{equation}
\delta_{CP}^{\rm test}/\pi \in [-0.554,\,-0.427],\quad
\varphi_{21}^{\rm test}/\pi \in [-0.745,\,-0.395]
\end{equation}
at $3\sigma$, corresponding to widths of $0.127$ and $0.350$ in units of $\pi$ respectively. The $3\sigma$ area is approximately $0.049\,\pi^2$. The large ${\chi^2}_{\mathrm{min}}^{\mathrm{DUNE}}=485.853$ confirms that DUNE's matter-enhanced oscillations at approximately $L=1300$~km provide strong rejection of the wrong mass ordering even after marginalization over non-unitary parameters. The elongation of the contours along the $\varphi_{21}$ direction (width $0.350\pi$) relative to the $\delta_{CP}$ direction (width $0.127\pi$) reveals a residual correlation: when the wrong mass ordering is assumed, the test hypothesis can partially compensate by adjusting $\varphi_{21}$ away from zero, generating a non-unitary-induced fake CP contribution that partially mimics the genuine phase.
\\
\\
\textbf{Hyper-Kamiokande alone} (panel b). The allowed region is dramatically larger. At $1\sigma$ it already spans $\delta_{CP}^{\rm test}/\pi\in[-0.650,\,-0.363]$ and $\varphi_{21}^{\rm test}/\pi\in[-0.236,\,+0.082]$, with area $0.072\,\pi^2$ -- already 7.1× larger than the combined result at the same confidence level. At $2\sigma$ and $3\sigma$ the $\varphi_{21}$ range expands to essentially the full parameter space, $\varphi_{21}^{\rm test}/\pi\in[-1.000,\,+0.974]$, while $\delta_{CP}^{\rm test}/\pi$ remains confined near the true value ($[-0.777,\,-0.172]$ at $3\sigma$). The $3\sigma$ area of $0.960\,\pi^2$ is 16.3× larger than the combined result. This dramatic vertical structure in the $\varphi_{21}$ direction is the direct signature of the hierarchy--CP--non-unitarity degeneracy of Equation~\eqref{eq:hcn} : at $L=295$~km, matter effects are too small to constrain $\varphi_{21}^{\rm test}$ through the wrong-ordering test, so virtually any value of $\varphi_{21}^{\rm test}$ can accommodate the observed event rates equally well at a fixed $\delta_{CP}^{\rm test}$ near the true value. The small ${\chi^2}_{\mathrm{min}}^{\mathrm{HK}}=4.404$, compared to DUNE's 485.853, quantifies this directly : Hyper-Kamiokande has very limited power to exclude the wrong mass ordering at this baseline, since the oscillation probability differences between NO and IO are too small without the matter-effect lever arm.
\\
\\
\textbf{Combined DUNE + Hyper-Kamiokande} (panel c). The combination breaks the degeneracy almost completely. At $1\sigma$ the combined allowed region has area $0.010\,\pi^2$ -- identical to DUNE alone, confirming that at this confidence level DUNE provides the dominant constraint. At $2\sigma$ and $3\sigma$ the combined area ($0.029\,\pi^2$ and $0.059\,\pi^2$ respectively) remains within a factor of $0.8$--$1.0$ of DUNE alone, while being 15.8× and 16.3× smaller than Hyper-Kamiokande alone at the same levels. The $\varphi_{21}$ range at $3\sigma$ is confined to $[-0.745,\,-0.363]$ which is a width of only $0.382\pi$ compared to Hyper-Kamiokande's full-range $1.974\pi$. 
\\
It is notable that the combined $3\sigma$ area ($0.059\,\pi^2$) is marginally larger than DUNE's alone ($0.049\,\pi^2$). This is physically expected rather than paradoxical : the combined $\chi^2$ must be consistent with both experiments simultaneously, and the specific $(\delta_{CP},\varphi_{21})$ combinations that Hyper-Kamiokande cannot exclude (the vertical band in panel b) partially overlap with the edge of DUNE's own allowed region, adding a few grid points at the contour boundary. The dominant constraint still comes from DUNE, but Hyper-Kamiokande's contribution is not purely additive, it modifies the contour shape slightly at $2\sigma$--$3\sigma$ by selecting only those parameter combinations consistent with both experiments' event rates.
\\
The improvement from the combination is therefore not simply a consequence of increased event statistics, but originates from the complementary degeneracy structures of the two experiments : the wide band of $\varphi_{21}^{\rm test}$ values that Hyper-Kamiokande cannot exclude corresponds precisely to the region that DUNE excludes at high confidence through its matter-enhanced oscillation pattern. Their intersection in parameter space is far smaller than either individual allowed region, and the combined analysis selects only those $(\delta_{CP}^{\rm test},\varphi_{21}^{\rm test})$ combinations consistent with both experiments simultaneously.

\section{Complementarity of DUNE and Hyper-Kamiokande}

The degeneracy analysis of the previous section demonstrates that neither DUNE nor Hyper-Kamiokande can independently resolve the hierarchy--CP--non-unitarity degeneracy, but their combination does so effectively. The physical origin of this complementarity lies in the fundamentally different roles that matter effects play at the two baselines.
\\
At DUNE ($L = 1300$~km), the charged-current matter potential $V_{cc}\approx 1.01\times10^{-22}$~GeV accumulated over the long baseline produces a significant MSW effect that depends strongly on the mass ordering. Both $\delta_{CP}$ and $\varphi_{21}$ enter the oscillation probability through matter-modified amplitudes, 
introducing correlations between them. This is reflected in DUNE's contours being elongated along the $\varphi_{21}$ axis (width $0.350\pi$ vs $0.127\pi$ in $\delta_{CP}$ at $3\sigma$) -- DUNE constrains $\delta_{CP}$ more tightly than $\varphi_{21}$ because $\delta_{CP}$ enters primarily through the matter-enhanced appearance amplitude, while $\varphi_{21}$ enters through a combination of matter-modified terms that is less tightly constrained by the integrated event rates alone.
\\
At Hyper-Kamiokande ($L=295$~km), matter effects are negligible and the oscillation probability is close to its vacuum form. The sensitivity to $\delta_{CP}$ arises through the standard vacuum interference term proportional to $\sin\delta_{CP}\sin\theta_{13}\sin\theta_{23}\sin(\Delta m^2_{31}L/4E)$, while $\varphi_{21}$ enters through a different combination of vacuum amplitudes involving the off-diagonal element $\alpha_{21}e^{i\varphi_{21}}$ of the non-unitary matrix. Without the matter-induced mixing of these contributions, Hyper-Kamiokande's sensitivity to $\varphi_{21}$ through the wrong-ordering test is essentially zero at $2\sigma$ and beyond (full $\varphi_{21}$ range allowed), while its sensitivity to $\delta_{CP}$ remains ($3\sigma$ width $0.700\pi$). This asymmetry is directly visible in the vertical band structure of panel~(b) of Figure.~\ref{fig:Degeneracy}.
\\
The combination exploits this asymmetry precisely. DUNE's $\chi^2$ surface has a deep minimum in the $\delta_{CP}$ direction but a broader valley in $\varphi_{21}$. Hyper-Kamiokande's $\chi^2$ surface is nearly flat in $\varphi_{21}$ but rises steeply away from $\delta_{CP}^{\rm true}$ even at this short baseline. The combined $\chi^2=\chi^2_{\rm DUNE}+\chi^2_{\rm HK}$ inherits the $\varphi_{21}$ constraint primarily from DUNE and the confirmation of $\delta_{CP}$ from Hyper-Kamiokande, producing an allowed region that is $16.3\times$ smaller than Hyper-Kamiokande alone at $3\sigma$.
\\
The hierarchy-discrimination power of the combined analysis can be further characterized through
\begin{equation}
\mathcal{H} = \chi^2_{\rm IO} - \chi^2_{\rm NO},
\end{equation}
which measures the preference for the true ordering over the wrong one. The combined minimum ${\chi^2}_{\mathrm{min}}^{\mathrm{combined}}=507.867$ is significantly larger than HK's minimum of $4.404$ and comparable to DUNE's $485.853$, confirms that the wrong mass ordering remains strongly excluded by the joint dataset, and that the non-unitary parameters provide insufficient additional freedom to mimic the true NO event spectra at IO when both experiments must be fitted simultaneously.
\\
Together, these results establish the combined DUNE and Hyper-Kamiokande program as a uniquely powerful strategy for simultaneously determining the neutrino mass ordering, the genuine Dirac CP phase $\delta_{CP}$, and the non-unitary CP phase $\varphi_{21}$ in the presence of the three-way hierarchy--CP--non-unitarity degeneracy. The complementarity demonstrated here is structural rather than merely statistical; it arises from the qualitatively different sensitivity of the two 
experiments to the matter potential, and persists as long as their baselines remain sufficiently different to probe distinct oscillation regimes.

\section{CP Violation Discovery Potential}

\begin{figure}[htbp]
\centering
\begin{subfigure}[t]{0.49\textwidth}
    \centering
    \includegraphics[width=\linewidth]{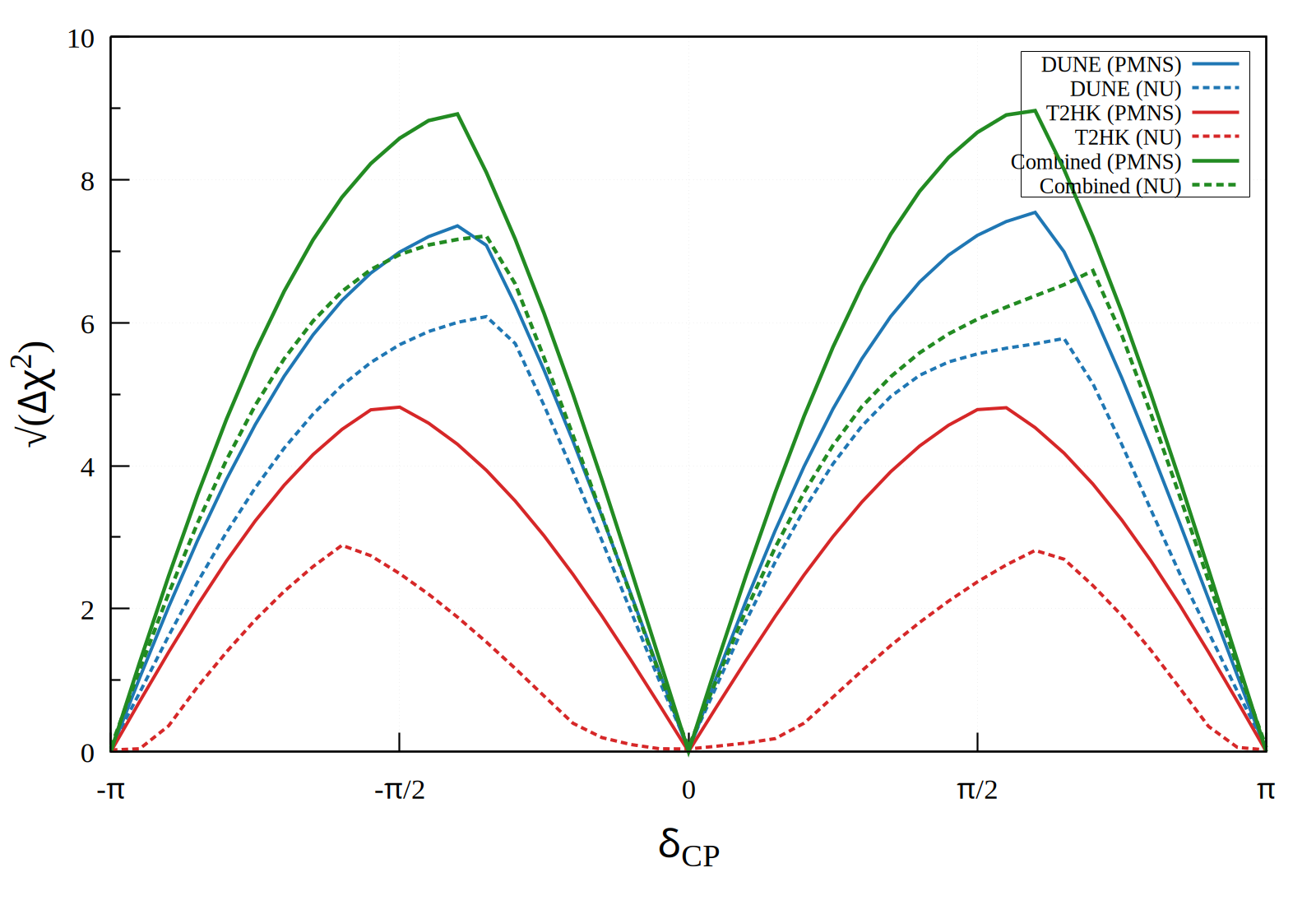}
    \caption{PMNS and non-unitarity (NU) CP violation discovery sensitivities.}
    \label{fig:CPV_NU_PMNS}
\end{subfigure}
\hfill
\begin{subfigure}[t]{0.49\textwidth}
    \centering
    \includegraphics[width=\linewidth]{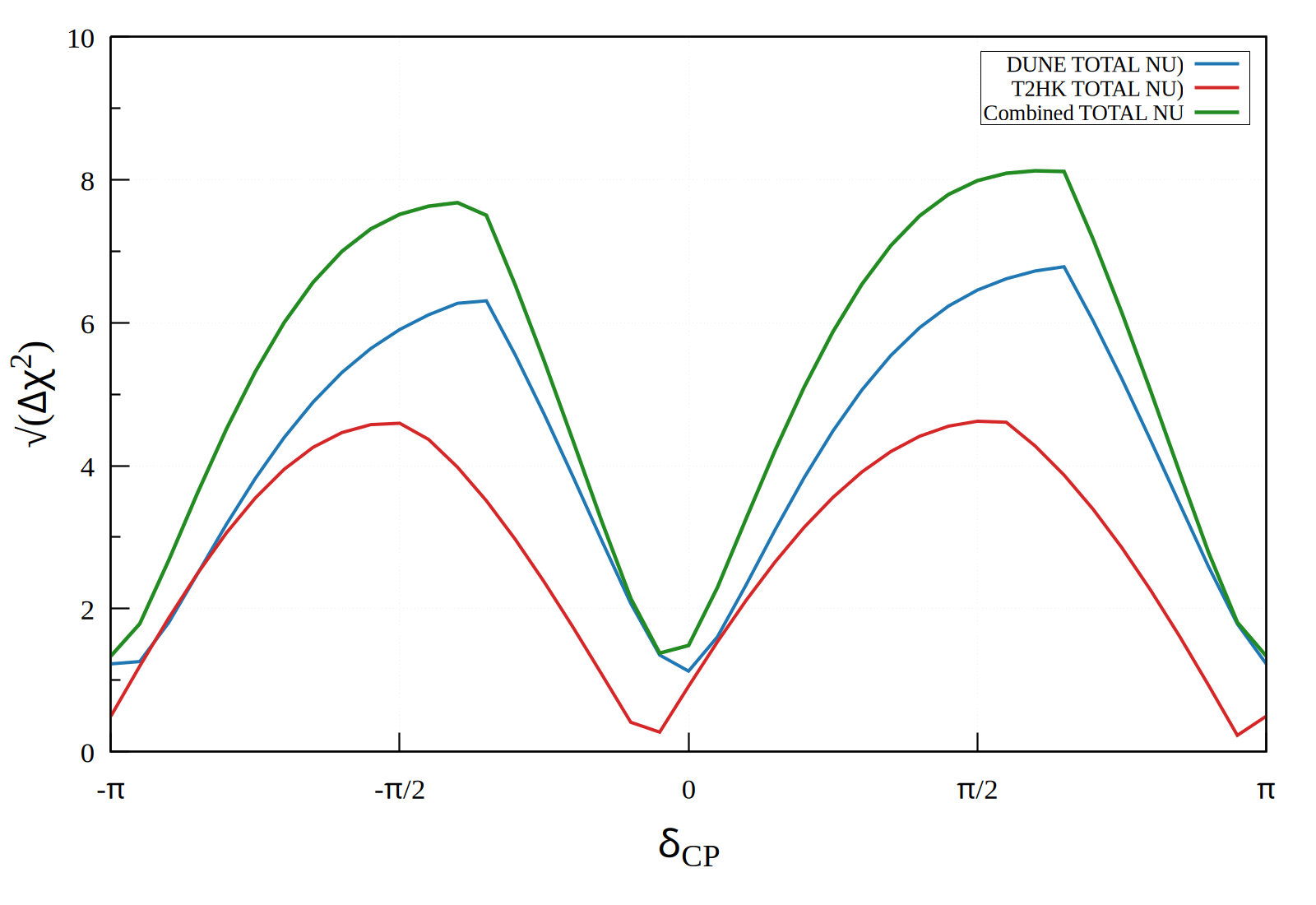}
    \caption{Total CP violation discovery sensitivity.}
    \label{fig:CPV_TOTAL}
\end{subfigure}
\caption{
Comparison of the CP violation discovery sensitivity for DUNE, Hyper-Kamiokande, and their combined analysis. Panel (a) shows the separate contributions from the standard PMNS CP phase (solid curves) and the non-unitarity (NU) CP phase (dotted curves), expressed as $\sqrt{\Delta\chi^2}$ as a function of the true CP phase $\delta_{CP}$. Panel (b) shows the total CP violation discovery sensitivity obtained from the simultaneous contributions of the PMNS and NU CP-violating phases. The sensitivities are calculated after marginalizing over the relevant oscillation and non-unitary parameters.
}
\label{fig:CPV_comparison}
\end{figure}

The CP-violation discovery sensitivity for DUNE, Hyper-Kamiokande, and their combined analysis is presented in Figure.~\ref{fig:CPV_comparison}. Figure~\ref{fig:CPV_NU_PMNS} compares the discovery reach obtained in the standard PMNS framework with that in the presence of non-unitary leptonic mixing, while Figure.~\ref{fig:CPV_TOTAL} shows the corresponding total CP-discovery sensitivity after including both the standard and non-unitary contributions.
As expected, the discovery sensitivity vanishes at the CP-conserving values $\delta_{CP}=0$ and $\pm\pi$, where the oscillation probabilities exhibit no intrinsic CP violation. The sensitivity increases as $|\delta_{CP}|$ approaches $\pi/2$, where the interference term proportional to $\sin\delta_{CP}$ becomes maximal, producing the largest separation between the CP-conserving and CP-violating hypotheses.
\\
Figure~\ref{fig:CPV_NU_PMNS} illustrates the impact of non-unitary leptonic mixing on the CP-discovery potential of the individual experiments. In the standard PMNS framework (solid curves), the combined DUNE+Hyper-Kamiokande configuration achieves the highest sensitivity throughout the entire $\delta_{CP}$ range, followed by DUNE and Hyper-Kamiokande. This behaviour reflects the complementarity between the long-baseline configuration of DUNE, which experiences strong matter effects, and the shorter-baseline Hyper-Kamiokande experiment, whose oscillation probabilities remain close to vacuum. Their combination simultaneously exploits matter-induced enhancement and high-statistics measurements, thereby providing the strongest sensitivity to leptonic CP violation.
\\
The dashed curves represent the corresponding sensitivities in the presence of non-unitary leptonic mixing. The additional non-unitary CP phase introduces new interference terms that partially reproduce the effects of the standard Dirac phase, leading to a systematic reduction of the CP-discovery reach. The reduction is particularly pronounced for Hyper-Kamiokande, whose peak sensitivity decreases from approximately $\sqrt{\Delta\chi^2}\simeq4.82$ to about $2.89$ which corresponds to nearly $65\%$ reduction. Since matter effects are relatively weak at the Hyper-Kamiokande baseline, the experiment possesses limited capability to distinguish genuine PMNS-induced CP violation from CP asymmetries generated by the additional non-unitary phase. Consequently, the hierarchy-CP-non-unitarity degeneracy becomes significantly stronger.
\\
In contrast, DUNE exhibits a considerably smaller degradation. Its peak sensitivity decreases from approximately $\sqrt{\Delta\chi^2}\simeq7.54$ to about $6.09$, corresponding to a reduction of nearly $35\%$. The stronger matter effects experienced over the $1300~\mathrm{km}$ baseline introduce additional information that constrains the non-unitary parameter space and suppresses a significant fraction of the fake CP solutions.
\\
Most importantly, the combined DUNE+Hyper-Kamiokande analysis retains a substantially larger CP-discovery reach than either experiment individually. Although non-unitary mixing reduces the peak sensitivity from approximately $\sqrt{\Delta\chi^2}\simeq8.97$ to about $7.21$, the degradation is considerably smaller than that observed for Hyper-Kamiokande alone. This demonstrates that the complementary baselines, neutrino energies, detector technologies, and matter effects of the two experiments effectively break the parameter degeneracies introduced by non-unitary mixing.
\\
Figure~\ref{fig:CPV_TOTAL} summarizes the total CP-discovery sensitivity after simultaneously accounting for both the standard and non-unitary CP-violating contributions. The combined analysis consistently provides the largest discovery significance over the entire $\delta_{CP}$ interval, reaching a maximum sensitivity of approximately $\sqrt{\Delta\chi^2}\simeq8$. DUNE remains the second most sensitive configuration, while Hyper-Kamiokande alone exhibits the lowest discovery reach because of its stronger susceptibility to non-unitarity-induced degeneracies. The smooth behaviour of the total sensitivity demonstrates that the combination of the two experiments efficiently recovers much of the sensitivity lost in the individual analyses.
\\
These results constitute one of the principal outcomes of this work. They demonstrate that although non-unitary leptonic mixing can significantly degrade the CP-discovery potential by introducing additional CP phases that mimic genuine leptonic CP violation, a joint analysis of DUNE and Hyper-Kamiokande largely overcomes these effects. The improvement originates from the complementary oscillation baselines and matter effects rather than simply increased statistics, providing a robust strategy for separating genuine PMNS-induced CP violation from fake CP violation generated by non-unitary leptonic mixing.
\section{Mass Hierarchy Discovery Sensitivity}

\begin{figure}[htbp]
\centering

\begin{subfigure}[t]{0.49\textwidth}
    \centering
    \includegraphics[width=\linewidth]{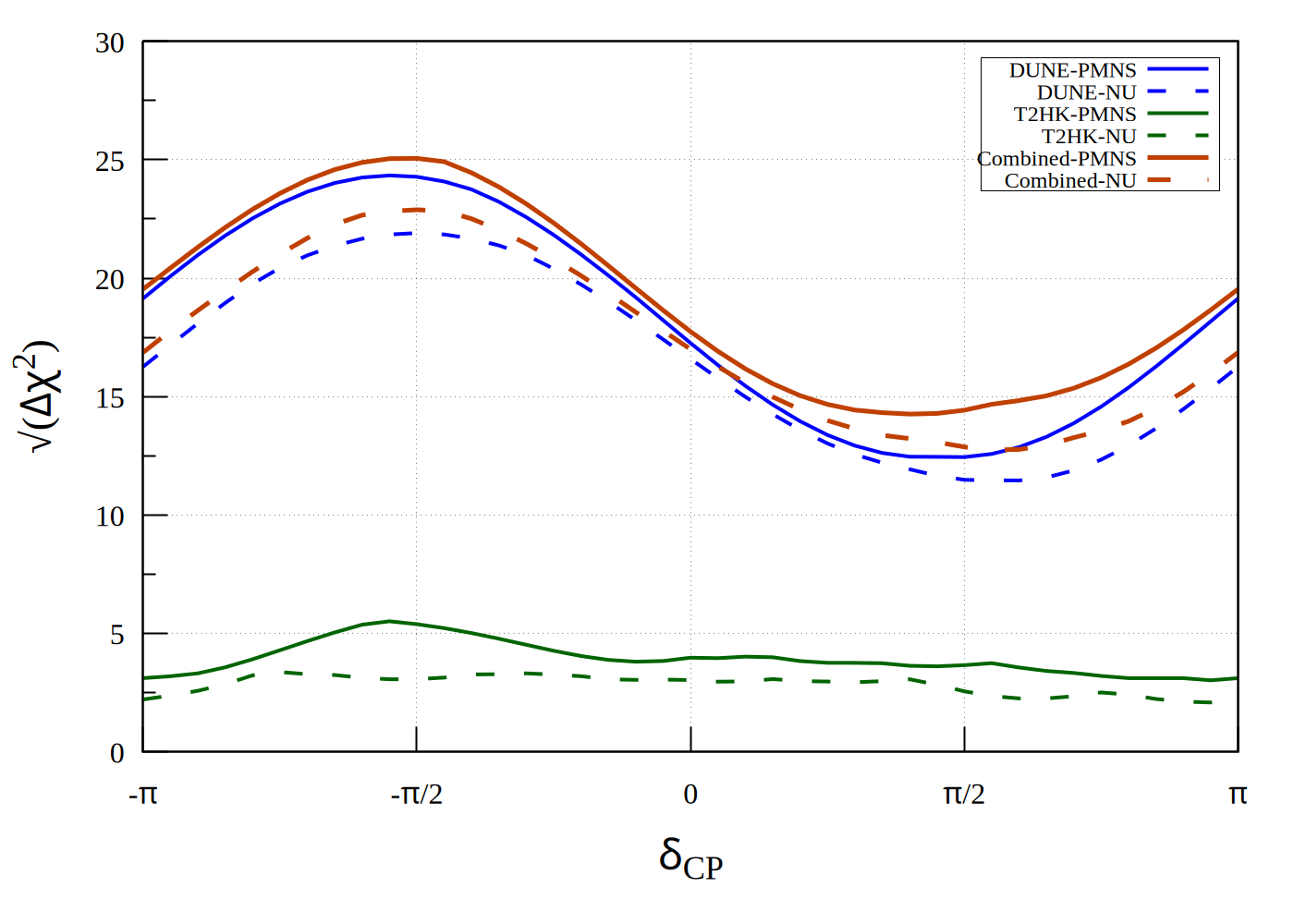}
    \caption{Comparison of the PMNS and NU contributions to the mass hierarchy discovery sensitivity.}
    \label{fig:MH_PMNS_NU}
\end{subfigure}
\hfill
\begin{subfigure}[t]{0.49\textwidth}
    \centering
    \includegraphics[width=\linewidth]{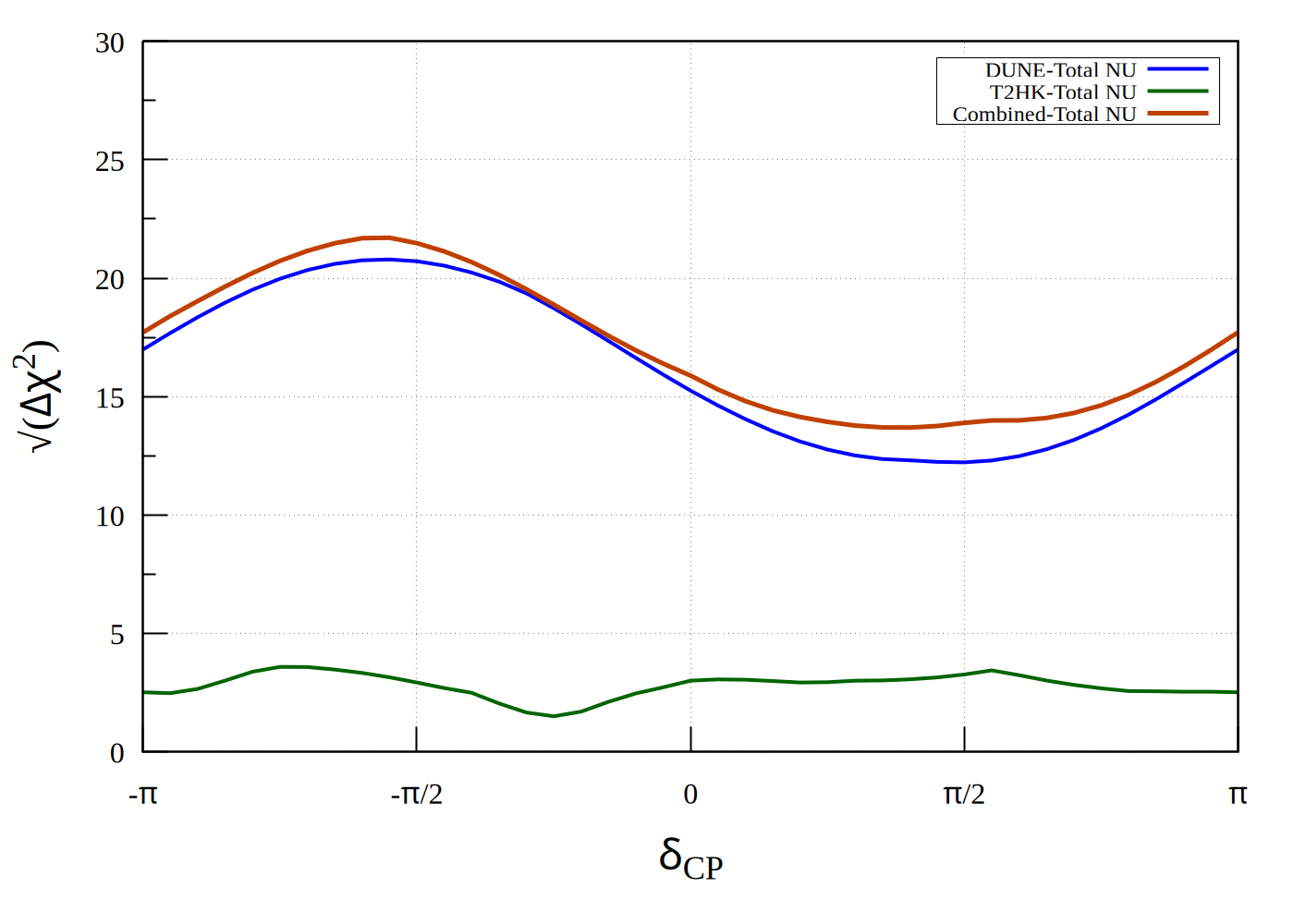}
    \caption{Total mass hierarchy discovery sensitivity including both PMNS and NU contributions.}
    \label{fig:MH_TOTAL}
\end{subfigure}

\caption{
Comparison of the neutrino mass hierarchy discovery sensitivity for DUNE, Hyper-Kamiokande, and their combined analysis as a function of the true CP-violating phase $\delta_{CP}$. Panel (a) compares the separate contributions from the standard PMNS framework and the NU scenario, while panel (b) shows the total mass hierarchy discovery sensitivity obtained after simultaneously including both PMNS and NU effects. In all cases, the sensitivities are expressed as $\sqrt{\Delta\chi^2}$ and are evaluated after marginalizing over the relevant oscillation and non-unitary parameters.
}
\label{fig:MH_comparison}

\end{figure}

The neutrino mass hierarchy discovery sensitivity for DUNE, Hyper-Kamiokande, and their combined analysis is presented in Figure.~\ref{fig:MH_comparison} as a function of the true Dirac CP phase $\delta_{CP}$. Figure~\ref{fig:MH_PMNS_NU} compares the hierarchy sensitivities obtained in the standard PMNS framework with those in the presence of non-unitary leptonic mixing, while Figure.~\ref{fig:MH_TOTAL} shows the corresponding total hierarchy sensitivity after simultaneously accounting for both the standard and non-unitary contributions.
\\
As shown in Figure.~\ref{fig:MH_PMNS_NU}, DUNE exhibits significantly larger hierarchy sensitivity than Hyper-Kamiokande throughout the entire $\delta_{CP}$ range. This behaviour is a direct consequence of the long $1300~\mathrm{km}$ baseline of DUNE, where matter effects strongly enhance the difference between the normal and inverted mass orderings. In contrast, the shorter $295~\mathrm{km}$ baseline of Hyper-Kamiokande experiences much weaker matter effects, resulting in a comparatively limited hierarchy discrimination. The combined DUNE+Hyper-Kamiokande analysis consistently provides the highest sensitivity over the full parameter space, demonstrating the complementarity of the two experiments.
\\
The dashed curves in Figure.~\ref{fig:MH_PMNS_NU} illustrate the effect of non-unitary leptonic mixing. The presence of the additional non-unitary phase introduces new correlations among the oscillation parameters and partially obscures the hierarchy-dependent matter effects. Consequently, the hierarchy sensitivity is reduced for all experimental configurations. The reduction is most pronounced for Hyper-Kamiokande, whereas DUNE remains comparatively robust owing to its stronger matter effects. Although the combined analysis also experiences a reduction, it retains the highest hierarchy sensitivity throughout the entire $\delta_{CP}$ interval, indicating that the complementary baselines substantially suppress the hierarchy--CP--non-unitarity degeneracy.
\\
Figure~\ref{fig:MH_TOTAL} presents the total hierarchy discovery sensitivity after including both the standard and non-unitary CP-violating contributions. The combined configuration achieves a maximum sensitivity of approximately $\sqrt{\Delta\chi^{2}}\simeq22$, followed closely by DUNE with $\sqrt{\Delta\chi^{2}}\simeq21$, while Hyper-Kamiokande reaches only about $\sqrt{\Delta\chi^{2}}\simeq3.5$. Unlike the CP-discovery sensitivity, the hierarchy significance remains non-zero throughout the entire $\delta_{CP}$ range because matter effects provide an intrinsic distinction between the two mass orderings even at CP-conserving values of $\delta_{CP}$. Although the hierarchy sensitivity exhibits a moderate dependence on $\delta_{CP}$ through the interference between matter and CP-violating terms, the ordering discrimination continues to be dominated by the matter potential, particularly for DUNE and the combined analysis.
\\
Overall, the results demonstrate that non-unitary mixing degrades the hierarchy discovery potential by introducing additional parameter degeneracies. Nevertheless, the combined DUNE+Hyper-Kamiokande analysis preserves most of the hierarchy sensitivity because the different baselines and oscillation energies probe complementary regions of the oscillation parameter space. This complementarity effectively constrains the additional non-unitary parameters and significantly improves the robustness of the mass-ordering determination compared with either experiment individually.

\begin{table}[ht]
\centering
\caption{Maximum discovery sensitivities for CP violation and mass hierarchy in different scenarios for DUNE, Hyper-Kamiokande (HK), and the combined DUNE+HK configuration.}
\label{tab:summary_sensitivity}
\begin{tabular}{lccc}
\toprule
\textbf{Observable} & \textbf{DUNE} & \textbf{HK} & \textbf{DUNE+HK} \\
\midrule
CPV (PMNS)      & 7.54  & 4.82 & 8.97 \\
CPV (NU)        & 6.09  & 2.89 & 7.21 \\
CPV (Total NU)  & 6.78  & 4.62 & 8.13 \\
\midrule
MH (PMNS)       & 24.34 & 5.51 & 25.06 \\
MH (NU)         & 21.90 & 3.36 & 22.89 \\
MH (Total NU)   & 20.79 & 3.58 & 21.71 \\
\bottomrule
\end{tabular}
\end{table}

\section{Conclusions}

In this work, we have investigated the impact of non-unitary leptonic mixing on the determination of leptonic CP violation and neutrino mass ordering in future long-baseline neutrino oscillation experiments. Working within the minimal non-unitarity framework, we examined how the additional CP-violating phase associated with the parameter $\alpha_{21}$ modifies neutrino oscillation probabilities and generates fake CP-violating signatures that can mimic the effects of the standard Dirac phase $\delta_{CP}$.
\\
Our results demonstrate that non-unitarity introduces significant hierarchy--CP parameter degeneracies, leading to a reduction in both CP-violation discovery and mass ordering sensitivities for individual experiments. The degradation is found to be considerably stronger for Hyper-Kamiokande, where matter effects are relatively weak, while DUNE remains more robust because of its long baseline and enhanced matter effects. The additional non-unitary phase produces observable CP asymmetries even for the CP-conserving values $\delta_{CP}=0$ and $\pi$, illustrating that an observed CP asymmetry cannot, by itself, be regarded as unambiguous evidence for genuine PMNS-induced CP violation.
\\
A central result of this work is that the combination of DUNE and Hyper-Kamiokande substantially alleviates these degeneracies. The complementary baselines, neutrino energies, detector technologies, and sensitivities to terrestrial matter effects significantly improve the simultaneous determination of the standard Dirac phase, the neutrino mass ordering, and the non-unitary parameters. In particular, the combined analysis considerably reduces the allowed $(\delta_{CP},\phi_{21})$ parameter space relative to the individual experiments, thereby suppressing fake CP solutions arising from non-unitary mixing.
\\
These results establish that future precision measurements of leptonic CP violation should be interpreted within a framework that allows for possible deviations from unitarity. The strategy developed in this work demonstrates that the combined analysis of long-baseline experiments provides a powerful and robust approach for disentangling genuine leptonic CP violation from new-physics effects associated with heavy neutral leptons. Such studies will play an important role in testing the unitarity of the leptonic mixing matrix and in probing physics beyond the Standard Model in the precision neutrino era.
\bibliographystyle{unsrt}
\bibliography{references}
\end{document}